\documentclass[
 reprint,
 superscriptaddress,
 amsmath,amssymb,
 aps,
pra,
]{revtex4-2}
\usepackage{amsmath,amssymb,amsthm}
\usepackage{graphicx}
\usepackage{booktabs}
\usepackage[hidelinks,colorlinks=true,linkcolor=teal, citecolor=blue, bookmarks=true,breaklinks=true]{hyperref}
\usepackage{enumitem}
\usepackage{xcolor}
\usepackage[caption=false]{subfig}
\usepackage{multirow}
\usepackage{siunitx}
\usepackage{physics}
\usepackage[normalem]{ulem}
\usepackage{cancel}

\hypersetup{
  colorlinks=true,
  linkcolor=blue,
  urlcolor=black,
  citecolor=blue
}

\newcommand{\R}{\mathbb{R}}

\definecolor{zwpurple}{RGB}{128,0,128}

\long\def\comment#1{}

\newcommand{\pcsdpnl}{Physical and Computational Sciences Division, Pacific Northwest National Laboratory, Richland, WA, 99354, USA}
\newcommand{\eedpnl}{Energy and Environment Division, Pacific Northwest National Laboratory, Richland, WA, 99354, USA}
\newcommand{\uw}{Department of Electrical and Computer Engineering, University of Washington, Seattle, WA, 98195, USA}
\newcommand{\ubc}{Department of Electrical and Computer Engineering, the University of British Columbia, Vancouver, BC V6T 1Z4, Canada}
\newcommand{\pitt}{Department of Computer Science, University of Pittsburgh, Pittsburgh, PA, 15260, USA}

\begin{document}
\title{Benchmarking and Resource Analysis for Augmented-Lagrangian Quantum Hamiltonian Descent}
\author{Zeguan Wu}
\email{zew79@pitt.edu}
\affiliation{\pcsdpnl}
\affiliation{\pitt}

\author{Mingze Li}
\affiliation{\eedpnl}

\author{Muqing Zheng}
\affiliation{\pcsdpnl}

\author{Meng Wang}
\affiliation{\ubc}

\author{Junyu Liu}
\affiliation{\pitt}

\author{Samuel Stein}
\affiliation{\pcsdpnl}

\author{Ang Li}
\affiliation{\uw}

\author{Yousu Chen}
\affiliation{\eedpnl}

\author{Chenxu Liu}
\email{chenxu.liu@pnnl.gov}
\affiliation{\pcsdpnl}

\date{\today}

\begin{abstract}
Quantum Hamiltonian Descent (QHD) is a continuous optimization algorithm based on simulating a time-dependent quantum Hamiltonian whose potential energy encodes the objective function and whose kinetic energy promotes exploration through quantum interference and tunneling. While QHD is formulated for unconstrained optimization, many real-world optimization problems are constrained and highly nonconvex. In this paper, we benchmark AL-QHD, a hybrid framework that embeds QHD within the Augmented Lagrangian Method (ALM), thereby solving a sequence of unconstrained subproblems while using ALM to enforce constraints. We evaluate AL-QHD on standard nonconvex test functions and use iterative refinement to improve solution accuracy at fixed per-run qubit cost.
We also perform a gate-based resource analysis on ACOPF-derived power-system subproblems constructed from power-network data to estimate the quantum-computer scale required for practical applications.
Resource estimates on Texas7k-derived ACOPF instances show steep hard-gate scaling, reaching $\sim 4.46\times 10^7$ entangling gates in a NISQ-oriented model and $\sim 9.42\times 10^8$ T gates in a fault-tolerant model at $\sim 5.3\times 10^2$ active variables. These results suggest that AL-QHD is a useful framework for studying constrained nonconvex optimization with QHD, but that practical ACOPF-scale applications would likely require large-scale fault-tolerant quantum hardware.
\end{abstract}

\maketitle

\section{Introduction}
Quantum computing has drawn a significant amount of attention recently because of its promising speedups compared to classical computing~\cite{abbas2024challenges, shor1999polynomial, gisin2002quantum, farhi2014quantum, han2002quantum, georgescu2014quantum, kandala2017hardware, rebentrost2018quantum, woerner2019quantum}. Starting from Deutsch's theory \cite{deutsch1992rapid}, a large number of
quantum algorithms have been proposed, including Shor's algorithm for integer factorization~\cite{ shor1999polynomial}, Grover's algorithm
for unstructured search~\cite{grover1996fast}, quantum algorithms for Hamiltonian simulation~\cite{georgescu2014quantum, kandala2017hardware}, and quantum optimization algorithms \cite{shor1994algorithms,grover1996fast,lloyd1996universal,abbas2024challenges}.
Due to the wide applications of mathematical optimization in science and engineering, it is natural to design and develop quantum optimization algorithms. Existing attempts include algorithms that accelerate subroutines
of classical optimization methods (e.g., by using quantum linear system algorithms) \cite{kerenidis2020quantumipm,wu2025quantumdual,apers2023quantum}, and algorithms that are
developed to utilize properties of quantum mechanics such as superposition, interference, and tunneling \cite{farhi2014quantum}.

Quantum annealing~\cite{Das2008_annealing, Tasseff2024_annealing, kim2025quantum}, adiabatic quantum optimization~\cite{Albash2018_aqc, Bombieri2025_qao}, and the Quantum Approximate Optimization Algorithm (QAOA)~\cite{farhi2014quantum, Zhou2020_qaoa} are representative Hamiltonian-based approaches for quantum optimization. They are most naturally formulated for discrete binary variables: the objective is encoded as an Ising-type Hamiltonian and the algorithm searches for low-energy bit strings~\cite{farhi2014quantum,abbas2024challenges}. Applying these methods to continuous-variable optimization therefore requires first discretizing each variable onto a finite grid and then encoding the resulting grid values into binary or spin variables, for example, using fixed-point binary, one-hot/unary, domain-wall, or related encodings, which introduces additional qubit overhead and a resolution-accuracy tradeoff~\cite{zaborniak2024discretequadraticmodelqubo, Marchand2019, Chancellor_2019}.
Other quantum optimization methods, such as quantum interior-point and linear-system-based algorithms, instead seek to accelerate algebraic subroutines in structured convex optimization~\cite{kerenidis2020quantumipm,wu2025quantumdual,apers2023quantum}.

Recently, Quantum Hamiltonian Descent (QHD) has been proposed as a quantum dynamical algorithm natively designed for unconstrained continuous-variable optimization problems~\cite{leng2023quantum, kushnir2025qhdopt}. QHD can be viewed as a quantum analog of gradient descent in the sense that it is governed by a differential equation. However, QHD is fundamentally different from its classical counterpart. Gradient descent and its variants compute gradients at a given location and move along a descent direction; by making the step size infinitely small, one obtains a trajectory governed by an ordinary differential equation. With an initial point and the governing equation, the classical system moves to a point with (approximately) zero gradient, which can be a local minimizer in a nonconvex landscape.
In contrast, QHD evolves a wavefunction
over the search domain, where the objective function is encoded as the potential energy of a time-dependent
Hamiltonian and a kinetic term induce exploration. Under suitable assumptions on the objective function and
with properly chosen damping schedules, QHD is shown to converge to a global minimizer when the evolution time is sufficiently large \cite{leng2023quantum}.

However, many practical optimization problems involve constraints that must be satisfied. Examples include portfolio optimization with budget and risk constraints~\cite{markowitz1952portfolio,cornuejols2007optimization}, trajectory planning with collision-avoidance and dynamics constraints~\cite{lavalle2006planning,betts2010practical}, and machine-learning model fitting with fairness, sparsity, or robustness constraints~\cite{zafar2017fairness,tibshirani1996regression,xu2009robustness}. This is especially
true in power system operation and planning, where constraints arise from physical laws, including AC power flow,
and engineering limits, including generator and voltage bounds. Power system optimization problems such as AC optimal power flow (ACOPF)~\cite{aravena2023recent,holzer2024grid,low2014convex,molzahn2019survey} are typically formulated as nonlinear programs with nonconvex equality constraints containing bilinear voltage products and trigonometric functions. These problems are well known to admit many locally optimal operating points and can be challenging for local
optimization methods \cite{nocedal1999numerical}.

QHD is originally designed for unconstrained optimization. A straightforward way to apply an unconstrained solver to a constrained problem is to transform constraints into penalty terms. Strong penalties can, in principle, enforce feasibility and optimality, but they often create steep and poorly conditioned landscapes that are sensitive to approximation error, discretization resolution, and finite simulation accuracy. A widely used classical framework for reducing this difficulty is the augmented Lagrangian method (ALM), which combines Lagrange multiplier updates with moderate quadratic penalties and solves a sequence of unconstrained subproblems whose solutions are gradually driven toward feasibility. This suggests a natural hybrid strategy, i.e., embed QHD within an outer ALM loop and use QHD to approximately solve the unconstrained subproblem at each iteration. However, the effectiveness of this strategy, which we term ``AL-QHD'', is not guaranteed. Moreover, each ALM iteration modifies the QHD potential through multiplier and quadratic-penalty terms, which can add diagonal Hamiltonian terms after encoding.

This construction also leads to an important resource question. In a digital implementation of QHD, the continuous search domain must be represented with finite precision, the Hamiltonian evolution must be simulated, and both kinetic and potential evolutions must be implemented to sufficient accuracy~\cite{leng2023quantum}. While analog or quantum-annealing implementations may provide useful heuristic or near-term demonstrations, they offer less direct control over coherent control errors, Hamiltonian miscalibration, decoherence, and thermal noise, which can be difficult to correct systematically~\cite{Young2013, MartinMayor2015, Albash_2019, Pearson2019}. If QHD is to deliver high-accuracy optimization solutions for nontrivial practical instances, a digital gate-based implementation with error-correction capability may ultimately be required. The digital circuit model also provides a concrete framework for counting qubits, gates, and other costly operations. 
The constrained setting further increases the cost, because each ALM iteration modifies the potential through multiplier and quadratic-penalty terms, which can introduce additional terms after Hamiltonian encoding.

In this work, we empirically assess the AL-QHD framework on representative constrained optimization examples and perform a gate-level resource analysis for digital implementations on practical optimization problems. Our goal is therefore to evaluate the feasibility and cost of AL-QHD on practical application cases, rather than the superiority of the QHD over classical optimization methods.
We firstly evaluate the solution quality of QHD for unconstrained optimization problems by classically simulating QHD via the split-step Fourier (pseudo-spectral) method. We observe that QHD can converge to global optimal solutions when the simulation is sufficiently accurate, and that iterative mesh refinement can dramatically improve solution quality without increasing the per-iteration cost.
We then evaluate the AL-QHD framework on constrained nonconvex problems, showing on a controlled nonlinear constrained benchmark that QHD's global-search behavior can support the ALM outer loop in finding a near-optimal feasible point in the correct basin under a moderate restart-budget comparison.
We further estimate the gate-level resources required to implement AL-QHD on digital quantum computers in both NISQ and fault-tolerant settings, with particular attention to the kinetic evolution, potential evolution, and additional ALM-induced terms. We find that applying AL-QHD to the AC optimal power flow problem at scales relevant for potential quantum advantage would likely require a large-scale fault-tolerant
quantum computer.

The rest of this paper is organized as follows.
Sec.~\ref{sec:background} reviews the QHD formulation and its discretization.
Sec.~\ref{sec:methodology} describes the benchmarking methodology, including the ALM framework, one-hot encoding, and gate counting model.
Sec.~\ref{sec:refinement} presents the iterative refinement procedure for improving low-resolution QHD solutions.
Sec.~\ref{sec:experiments} reports experimental results on standard test functions, including an unconstrained benchmark on the Ackley function and a constrained benchmark on the scaled Rastrigin function, and includes an ACOPF-derived power-system resource-estimation case study in Sec.~\ref{sec:power}.
Sec.~\ref{sec:discussion} discusses findings and implications, and Sec.~\ref{sec:conclusion} concludes our work.

\section{Background: Quantum Hamiltonian Descent}
\label{sec:background}

In this section, we briefly review the formulation of QHD for unconstrained
continuous optimization.
The purpose is to establish notation and highlight the key dynamical
ingredients of QHD that are relevant for benchmarking and implementation. For the sake of simplicity, we will avoid mentioning several technical assumptions for the convergence of QHD and refer readers to the original QHD paper~\cite{leng2023quantum}.

\subsection{Unconstrained optimization and QHD Hamiltonian}
We consider unconstrained optimization problems of the form
\begin{equation}
  \min_{x \in \R^d} f(x),
\end{equation}
where $f : \R^d \to \R$ is assumed to be continuously differentiable and possibly nonconvex, and $d$ is the dimension of the variable $x$.
In classical optimization, gradient-based methods generate a single trajectory in the decision space,
typically converging to a stationary point that may be only locally optimal \cite{nocedal1999numerical}.
QHD takes a fundamentally different approach.
Instead of evolving a single point, QHD evolves a quantum wavefunction $\Psi(x,t)$ over the
continuous domain, whose dynamics are governed by the Schr\"odinger equation
\begin{equation}
  i\,\frac{d}{dt}\Psi(t) = \widehat{H}(t)\,\Psi(t),
  \label{eq:schrodinger}
\end{equation}
where $\widehat{H}(t)$ is a time-dependent Hamiltonian operator.
The QHD Hamiltonian is defined as
\begin{equation}
  \widehat{H}(t) = a(t)\,\widehat{K} + b(t)\,\widehat{V},
  \quad
  \widehat{K} = -\frac{1}{2}\nabla^2,
  \quad
  \widehat{V} = f(x),
  \label{eq:qhd_hamiltonian}
\end{equation}
with $a(t) = e^{\phi(t)}$ and $b(t) = e^{\chi(t)}$ denoting prescribed time-dependent damping
functions.
Here, $\widehat{K}$ is the kinetic energy operator and $\widehat{V}$ is the potential energy operator
associated with the objective function.

From a physical perspective, QHD describes the motion of a quantum particle evolving in a potential
landscape defined by $f(x)$.
The kinetic term induces dispersion, interference, and tunneling effects, while the potential term biases the evolution toward regions of low objective value.
By choosing the damping schedules so that the relative weight of the kinetic energy decays over time, the wavefunction gradually concentrates near low-energy states.
Equivalently, the decaying kinetic prefactor can be interpreted as increasing the effective mass of the particle, which suppresses spatial spreading and increases localization of the wavefunction.
Under suitable assumptions on $f$ and the damping schedules, it can be shown that the probability
mass concentrates near global minimizers of $f$ as the evolution time increases \cite{leng2023quantum}.

The output of QHD is obtained by measuring the position of the particle, represented by the observable $x$, at the final time $T$. Repeated measurements yield samples distributed according to $|\Psi(x,T)|^2$, from which candidate
solutions to the optimization problem can be extracted.
In this sense, QHD can be viewed as a quantum generalization of gradient-based descent that explores the landscape through a distribution over trajectories rather than a single deterministic path.

This interpretation also connects QHD to annealing-based optimization. Both approaches use a time-dependent Hamiltonian whose early dynamics encourage exploration and whose later dynamics emphasize the objective landscape. In conventional quantum annealing or adiabatic optimization for Ising models, this is often achieved by reducing a transverse-field driver while increasing the problem Hamiltonian. In QHD, the analogous mechanism is implemented for continuous variables through the kinetic operator $-\nabla^2/2$ and the potential operator $f(x)$: the kinetic term plays the role of the driver that spreads the wavefunction, while the increasing relative weight of the potential term concentrates probability near low-objective regions. Thus QHD can be viewed as an annealing-like quantum dynamical method adapted to continuous optimization rather than binary Ising optimization.

\subsection{Time and space discretization}

To implement QHD on digital quantum hardware, the first-quantized dynamics must be discretized. The continuous search space is represented with finite precision, and the continuous-time Hamiltonian evolution is approximated by a sequence of discrete quantum operations. This discretized circuit representation allows us to estimate gate-level resources, such as qubit counts, gate counts, and rotation-synthesis costs, and provides a framework in which simulation errors can be systematically controlled in a fault-tolerant implementation. The same spatial and temporal discretization is also required for the classical simulations used in our benchmarking. We briefly describe the standard discretization procedure adopted in this work.

First, the continuous space domain $\R^d$ is truncated to a bounded region
$\Omega = [\ell, u]^d$ that is assumed to contain at least one global minimizer of the objective
function.
Each dimension is then discretized using a finite grid with resolution $r$, yielding a total of
$r^d$ grid points.
The discretization resolution controls the trade-off between solution accuracy and computational resources, both in classical simulation and in quantum implementation.

Second, the continuous-time Schr\"odinger evolution is approximated by Trotterization product formula.
Let $T$ denote the total evolution time and divide the interval $[0,T]$ into $M$ time steps of size
$\Delta t = T/M$.
The time-ordered evolution operator associated with the QHD Hamiltonian is approximated as
\begin{equation}
  U(T,0)
  \;\approx\;
  \prod_{j=0}^{M-1}
  \exp\!\left(-i\,\Delta t\,a(t_j)\,\widehat{K}\right)\,
  \exp\!\left(-i\,\Delta t\,b(t_j)\,\widehat{V}\right),
  \label{eq:trotter}
\end{equation}
where $t_j = j\,\Delta t$.
This first-order product formula separates the kinetic and potential evolutions, each of which has
a structure amenable to efficient implementation under appropriate encodings.
The kinetic operator $\widehat{K} = -\frac{1}{2}\nabla^2$ corresponds to a discrete Laplacian on the grid and induces transitions between neighboring grid points.
The potential operator $\widehat{V} = f(x)$ is diagonal in the position basis and can be implemented as a phase rotation conditioned on the grid value.
Higher-order product formulas can be used to reduce Trotter error, but for benchmarking purposes, we focus on the simplest decomposition.

In this paper, discretization serves two roles.
First, it enables classical simulation of QHD to assess solution quality under finite resolution.
Second, it provides a concrete basis for estimating the quantum resources required to implement the discretized Hamiltonian on gate-based quantum devices.
Our goal is not to optimize discretization schemes, but to use a transparent and reproducible setup
for comparative benchmarking.

\section{AL-QHD workflow for constrained optimization}
\label{sec:alm}
Many optimization problems of practical interest involve constraints, often in the form of
nonlinear equality and inequality conditions.
This is particularly true for engineering applications such as power system optimization~\cite{wood2013power,chen2022computing}, where
constraints arise from physical laws and operational limits.
To apply QHD to this type of constraint problems, it is natural to embed QHD within a framework that systematically handles constraints while preserving the ability to exploit QHD's global-search behavior.

We adopt ALM as the outer optimization framework and use QHD as an inner solver for the resulting sequence of unconstrained subproblems, which we term as ``AL-QHD'' workflow.
Before describing this hybrid workflow in detail, we first explain the motivation for this choice in Sec.~\ref{sec:alm:motivation}, and review the ALM construction in Sec.~\ref{sec:alm:alm}. We then discuss our AL-QHD workflow in Sec.~\ref{sec:alm:AL-QHD}.
\subsection{Motivation of ALM} \label{sec:alm:motivation}

Consider a constrained nonlinear program
\begin{equation}
  \min_{x\in\R^d} f(x)
  \quad \text{s.t.}\quad h_i(x)=0,\; i=1,\dots,m.
  \label{eq:nlp_eq}
\end{equation}
A straightforward approach to constrained optimization is to transform constraints into penalty terms and solve a single unconstrained problem.
For example, one may replace equality constraints $h_i(x)=0$ with a penalized objective of the form
$f(x) + \mu \sum_i h_i(x)^2$.
In classical optimization, such penalty methods are widely used, but their effectiveness depends critically on the choice of penalty parameters.
Large penalties can enforce feasibility but often lead to ill-conditioned landscapes that are
difficult for local solvers to navigate.

One might expect QHD, with its ability to explore nonconvex landscapes, to tolerate stronger penalty functions than classical gradient-based methods.
In principle, extremely strong or even ``exact'' penalties could be used to encode constraints
directly into the potential energy.
However, such penalties produce landscapes with very steep and narrow feasible
regions.
Under spatial discretization and approximate Hamiltonian simulation, these narrow regions can be easily distorted or eliminated, making the resulting problem highly sensitive to approximation
errors.

These considerations motivate a more conservative approach to constraint handling.
Rather than enforcing constraints through a single heavily penalized objective, we seek a framework
that introduces constraints gradually while maintaining moderate curvature in each subproblem.
ALM provides precisely this mechanism by combining Lagrange multipliers
with quadratic penalties and updating them iteratively.
This allows each unconstrained subproblem to remain well-shaped while driving the iterates toward
feasibility over successive iterations.

\subsection{Augmented Lagrangian Method} \label{sec:alm:alm}
ALM transforms Problem~\eqref{eq:nlp_eq} into a sequence of unconstrained subproblems by minimizing the augmented Lagrangian
\begin{equation}
  \mathcal{L}_\rho(x, \lambda)
  =
  f(x)
  + \sum_{i=1}^m \lambda_i h_i(x)
  + \frac{\rho}{2} \sum_{i=1}^m h_i(x)^2,
  \label{eq:alm}
\end{equation}
where $\lambda \in \R^m$ denotes the vector of Lagrange multipliers and $\rho > 0$ is a penalty
parameter.
Compared to simple quadratic penalty methods, the presence of the linear multiplier terms allows
ALM to enforce feasibility without requiring $\rho$ to grow excessively large.
Given multipliers $\lambda_k$ and penalty parameter $\rho_k$ at iteration $k$, the method
approximately solves the unconstrained subproblem
\begin{equation}
  x_{k+1}
  \;\approx\;
  \arg\min_x \mathcal{L}_{\rho_k}(x, \lambda_k),
\end{equation}
followed by an update of the multipliers
\begin{equation}
  \lambda_{k+1,i}
  =
  \lambda_{k,i}
  + \rho_k h_i(x_{k+1}),
  \qquad i = 1,\dots,m.
\end{equation}
The penalty parameter $\rho_k$ may be kept fixed or increased according to update rules \cite{nocedal1999numerical}.

From an algorithmic perspective, ALM separates constraint handling from unconstrained optimization.
Each iteration requires solving an unconstrained problem defined by $\mathcal{L}_{\rho_k}$, while
global feasibility is enforced across iterations through multiplier updates.
This structure makes ALM particularly suitable for hybrid approaches in which different solvers are used for the inner unconstrained subproblem.

\subsection{AL-QHD workflow} \label{sec:alm:AL-QHD}
We now describe the AL-QHD hybrid optimization workflow, where QHD is used as the inner solver within ALM.
At each ALM iteration, the constrained problem is reduced to an unconstrained subproblem defined
by the augmented Lagrangian, and QHD is applied to approximately minimize this subproblem.
Specifically, given the current multipliers $\lambda_k$ and penalty parameter $\rho_k$, the
augmented Lagrangian $\mathcal{L}_{\rho_k}(x,\lambda_k)$ defines an unconstrained objective function.
This objective is treated as the potential energy term in the QHD Hamiltonian, and the resulting quantum dynamics are simulated for a fixed evolution time.
The final wavefunction is measured to produce candidate solutions, from which a representative
minimizer $x_{k+1}$ is selected.
This point is then used to update the multipliers and, if necessary, the penalty parameter for the next ALM iteration.
The overall workflow can be summarized as follows:
\begin{enumerate}[leftmargin=*, noitemsep]
  \item Initialize multipliers $\lambda_0$ and penalty parameter $\rho_0$.
  \item For $k = 0,1,2,\dots,K_\text{alm}$, where $K_\text{alm}$ is the total number of ALM iterations:
  \begin{enumerate}[leftmargin=*, noitemsep]
    \item Form the augmented Lagrangian $\mathcal{L}_{\rho_k}(x,\lambda_k)$.
    \item Run QHD on the unconstrained objective $x \mapsto \mathcal{L}_{\rho_k}(x,\lambda_k)$
          to generate candidate solutions.
    \item Select an approximate minimizer $x_{k+1}$ from the QHD output.
    \item Update the multipliers $\lambda_{k+1}$ and, optionally, the penalty parameter $\rho_k$.
  \end{enumerate}
\end{enumerate}
In this hybrid approach, ALM is responsible for enforcing feasibility and coordinating progress toward constraint satisfaction, while QHD is used to address the nonconvexity of each
unconstrained subproblem.
This separation of roles allows QHD to operate in a regime where its global-search behavior is most relevant, without requiring it to directly encode hard constraints through extreme penalty terms.

\section{Benchmarking Methodology \& Resource Models}
\label{sec:methodology}
This section describes the benchmarking methodology used in this work and the associated resource
models for benchmarking QHD and for estimating quantum resources.
Our goal is not to propose an optimized encoding or compilation strategy, but to adopt a transparent and reproducible pipeline that allows consistent comparison across problem instances.
All implementation choices described here are applied uniformly throughout the experiments and
resource analyses reported in later sections.

\subsection{One-hot encoding and Ising Hamiltonian construction}
To estimate the quantum resources required to implement the AL-QHD workflow on quantum hardware, the continuous optimization problem must be encoded into a qubit Hamiltonian, similar to the QHD algorithm itself.
Several encodings can be used after discretizing continuous variables, including compact binary-type encodings, unary or one-hot encodings, and more general Hamiltonian-embedding constructions that trade Hilbert-space size against hardware-efficient dynamics~\cite{leng2025hamiltonianembedding}. In this work, we adopt and extend the one-hot encoding scheme used in \texttt{QHDOPT}~\cite{kushnir2025qhdopt}. The purpose is not to claim that one-hot encoding is uniquely optimal, but to exploit the analytical structure of the potential terms.

One-hot encoding uses $r$ qubits for a variable with $r$ grid points, rather than $O(\log r)$ qubits as in binary encodings, but it represents grid-value indicator functions directly as occupation operators. This larger qubit footprint gives more flexibility and leads to transparent, often lower-overhead constructions of diagonal potential-evolution gates for polynomial, trigonometric, and penalty terms.
Specifically, let $x_j$ be a continuous decision variable discretized into $r$ grid points
$\{g_{j,1}, g_{j,2}, \dots, g_{j,r}\}$.
We allocate $r$ qubits to represent $x_j$, where each basis state corresponds to exactly one active
qubit.
Within the one-hot subspace, the occupation operator associated with grid point $g_{j,k}$ can be
written as
\begin{equation}
  n_{j,k} = (I - Z_{j,k})/2,
\end{equation}
where $Z_{j,k}$ denotes the Pauli-$Z$ operator acting on the $k$-th qubit associated with variable
$x_j$.
This representation allows functions of a single variable to be encoded as diagonal operators.
For example, a function $f(x_j)$ is mapped to the operator
\begin{equation}
  f(x_j)
  \;\leadsto\;
  \sum_{k=1}^r f(g_{j,k})\, n_{j,k},
\end{equation}
which is diagonal in the computational basis.
Multivariate functions are constructed by taking products of the corresponding single-variable operators, resulting in sums of Pauli-$Z$ strings with varying locality.

\subsection{Resource analysis model}
\label{sec:meth:res_model}

The resource model differs substantially between NISQ and fault-tolerant implementations. In NISQ devices, where no quantum error correction is applied, resource costs are determined by the native physical gates supported by the hardware. Single-qubit gates are typically easier to implement than entangling gates, and phase rotations such as $R_z$ gates can often be realized efficiently through virtual frame updates or equivalent control-frame changes, depending on the platform~\cite{McKay2017}. By contrast, two-qubit gates require generating entanglement between physical qubits and usually dominate the circuit error rate, duration, and calibration overhead. For this reason, in our NISQ resource estimates, we treat two-qubit gates as the ``hard'' operations, while single-qubit gates are counted as comparatively inexpensive.

The cost model changes in the fault-tolerant quantum computing (FTQC) regime. Using the surface code as a representative example, logical Clifford operations, including Pauli operations, Hadamard gates, and CNOT or multi-qubit Pauli measurements implemented through lattice surgery, are usually regarded as relatively inexpensive compared with non-Clifford operations, although they still consume code cycles and logical qubit patches. By the Gottesman--Knill theorem, universal fault-tolerant computation requires operations beyond the Clifford group, while Clifford circuits alone are classically efficiently simulable~\cite{gottesman1998, Aaronson2004}. By the Eastin--Knill theorem, no quantum error-correcting code can implement a full universal gate set transversally~\cite{Eastin2009}. In surface-code architectures, the standard costly non-Clifford resource is the $T$ gate, typically supplied through magic-state distillation~\cite{Bombin2009, Horsman2012, Brown2017, Litinski2019, litinski2019magic}, cultivation~\cite{gidney2024cultivation, rosenfeld2025}, and code switching~\cite{Laflamme2014, Anderson2014}, and Clifford gates can be removed by merging them into multi-qubit Pauli rotations and measurements~\cite{Litinski2019}. Arbitrary rotations such as $R_z(\theta)$ are therefore not treated as native low-cost logical gates. Instead, they must be approximated by Clifford+$T$ sequences using synthesis methods such as Solovay--Kitaev~\cite{kitaev1997quantum, dawson2005solovay} or gridsynth~\cite{ross2014optimal} methods. Therefore, in our FTQC resource analysis, we treat Clifford operations as relatively ``easy'' logical operations and treat $T$ gates, especially those arising from decomposing $R_z$ rotation gates, as ``hard'' operations.

When counting the cost of the AL-QHD workflow, we use two complementary views:
\begin{itemize}[leftmargin=*, noitemsep]
  \item \textbf{Kinetic evolution cost:} extracted from compiled one-register kinetic circuits under the checked-in IonQ backend path and scaled linearly across identical one-hot registers. In the NISQ era, resource counting, we count the resulting native single- and two-qubit gates directly. This is equivalent to assuming NISQ implementation on IonQ experimental platforms.
  In the fault-tolerant implementation, we estimate the resource cost using a surface-code-based model. In this model, Clifford operations are treated as relatively low-cost logical operations that can be supported through lattice surgery, whereas non-Clifford operations are treated as the dominant hard resources. We therefore identify the non-Clifford operations in the corresponding NISQ circuit and decompose them into Clifford gates and $R_z$ rotations. For the kinetic evolution, we first count the $T$ and $T^\dagger$ gates that arise directly from $R_z$ rotations with angles $\pm \pi/4$ modulo $2\pi$. This count is denoted by $N_{T,K}$. We then count the remaining non-Clifford $R_z$ rotations, denoted by $N_{r,K}$, which must be synthesized into Clifford+$T$ sequences in the fault-tolerant setting.

  \item \textbf{Potential evolution cost:} estimated from the number and locality of Pauli-$Z$ strings using a parity decomposition. A $k$-local $Z$-string can be implemented with $2(k-1)$ CNOTs plus one $R_z(\theta)$ rotation (up to optimization), hence we estimate the number of Clifford gates by $2(\text{locality}(t)-1)$, and the number of $R_z$ rotations (noted as $N_{r,V}$) by the number of nontrivial Pauli-$Z$ rotation terms.
    Similar to the kinetic terms, in the fault-tolerant implementation, we treat the Clifford gates as easy gates, and $R_z$ gates as hard.

\end{itemize}
Note that in the fault-tolerant implementation, $R_z$ rotations are not treated as directly supported logical gates. Instead, each $R_z(\theta)$ rotation must be synthesized into a Clifford+$T$ sequence, using gates such as $H$, $S$, and $T$, to an approximation tolerance $\epsilon$. In the following resource estimate, we do not synthesize every rotation gate individually to construct the full Clifford+$T$ circuit. Instead, for each target tolerance $\epsilon$, we estimate a synthesis factor $r(\epsilon)$, defined as the average number of $T$ gates required to synthesize a single $R_z(\theta)$ rotation. We obtain $r(\epsilon)$ by sampling 5000 random rotation angles uniformly and averaging the resulting $T$ counts after synthesis using the \texttt{gridsynth} method~\cite{gridsynth}. Given this factor, the total fault-tolerant $T$ count is estimated as
\begin{equation}
    N_{t, \text{total}} = N_{t, K} +  r(\epsilon) \left( N_{r, K} + N_{r, V}\right).
\end{equation}
A more detailed discussion of how we determine the synthesis factor $r(\epsilon)$ can be found in Appendix~\ref{app:ft_factor}.

\section{Iterative refinement for low-resolution solutions}
\label{sec:refinement}

Spatial discretization is a major bottleneck for QHD on gate-based devices: increasing resolution improves accuracy but inflates qubit and gate counts. In many operational settings, however, a low-resolution solution that is \emph{good enough} can still be valuable, especially as a warm start for a classical solver. To reduce the qubit usage and to enable us to classically simulate the benchmark cases, we therefore use an \emph{iterative refinement} (adaptive zoom) procedure that achieves high accuracy from a coarse grid by progressively narrowing the search domain:

\begin{enumerate}[leftmargin=*, noitemsep]
  \item \textbf{Coarse QHD run:} choose a coarse resolution $r_0$ and broad bounds $\Omega_0$; run QHD to obtain samples.
  \item \textbf{Select promising region:} identify the peak of the final probability density $|\Psi(\mathbf{x}, T)|^2$ and define a
        smaller box $\Omega_1$ 
        from marginal probability intervals around the peak.
  \item \textbf{Refine:} rerun QHD on $\Omega_1$ with the same grid resolution $r_0$, effectively increasing the spatial resolution per unit length without increasing computational cost per step.
  \item \textbf{Iterate:} repeat the zoom procedure for a prescribed number of iterations $Z$ or until a convergence criterion is met.
  \item \textbf{Optional classical polish:} apply a local classical solver (e.g., L-BFGS-B or SLSQP)
        initialized at the best QHD sample. (In the experiments reported in this paper, no classical polish is applied; all reported solutions are raw QHD outputs.)
\end{enumerate}
More precisely, in Step 2, after zoom level $z$, let $p_z(\alpha)=|\Psi_z(\mathbf{x}_\alpha,T)|^2/\sum_\beta |\Psi_z(\mathbf{x}_\beta,T)|^2$ be the normalized probability on the grid. For each coordinate $j$, compute the one-dimensional marginal
\begin{equation}
  p_{z,j}(k)=\sum_{\alpha:\,\alpha_j=k} p_z(\alpha).
\end{equation}
Starting from the marginal peak $k_j^*=\arg\max_k p_{z,j}(k)$, we expand an index interval $I_j=[\ell_j,u_j]$ one grid point at a time, always adding the adjacent side with larger marginal probability, until
\begin{equation}
  \sum_{k=\ell_j}^{u_j} p_{z,j}(k)\ge \eta,
\end{equation}
where $\eta$ is a chosen probability threshold; in our AL-QHD workflow, we set $\eta=0.99$.
The next search box is the Cartesian product of the corresponding coordinate intervals,
\begin{equation}
  \Omega_{z+1}=\prod_{j=1}^d [x_{j,\ell_j},\,x_{j,u_j+1}],
\end{equation}
with endpoints clipped to the current domain. This construction preserves the high-probability region indicated by the QHD wavefunction while reducing the physical side length of the grid for the next run.

The key insight is that each zoom iteration reuses the same grid size, so the per-iteration cost (in time steps, qubits, and gates) remains constant. The effective resolution, however, improves geometrically with the number of zoom iterations. As we will demonstrate in Section~\ref{sec:experiments}, this mechanism allows QHD to push the solution accuracy far beyond the underlying coarse-grid spacing on low-dimensional test problems.

An important implementation detail is \emph{best-solution tracking}: across all zoom levels, the algorithm retains the best solution found so far. If a zoom iteration fails to improve the objective (e.g., due to Trotter error in a very narrow domain), the algorithm does not degrade.

We also note that the classical simulation of QHD via the split-step Fourier method is \emph{fully deterministic}: given the same grid, time discretization, and initial wavefunction, the evolution and the resulting probability density $|\Psi(\mathbf{x}, T)|^2$ are uniquely determined. The candidate solution is extracted as the grid point maximizing this density. Consequently, the results reported below are reproducible and do not require averaging over multiple runs or error bars.

\section{Experiments and evaluation}
\label{sec:experiments}
This section reports experimental results on two representative benchmark problems: an unconstrained optimization problem on the Ackley function and a constrained optimization problem on the scaled Rastrigin function. These two cases are designed to test complementary aspects of QHD. The Ackley function is essentially unimodal for gradient-based methods, where the entire domain forms a single basin of attraction around the global minimum, so the comparison with classical solvers isolates the \emph{precision} benefit of iterative refinement. The scaled Rastrigin function, by contrast, has approximately 900 local minima with narrow basins of attraction, so the comparison tests whether AL-QHD can recover the correct constrained basin under a budgeted multi-start classical baseline.

All QHD simulations use the QHD-C damping schedule from Ref.~\cite{leng2023quantum}:
\begin{equation}
  \widehat{H}(t) = \left(\frac{2}{s + t}\right)^3 \widehat{K} + 2t^3\, \widehat{V},
  \label{eq:qhd_c_schedule}
\end{equation}
where $s$ is a small regularization parameter. The Schr\"odinger evolution is simulated classically via the split-step Fourier (pseudo-spectral) method with Strang splitting, evolving from $t = 0$ to $t = 10$ in $N_t = 50{,}000$ time steps. The initial wavefunction is a uniform superposition over the computational domain, consistent with the QHD formulation.

\subsection{Unconstrained benchmark: Ackley function}
\label{sec:exp_ackley}
To benchmark the validity of the iterative refinement strategy, we focus on testing the QHD algorithm on an unconstrained optimization problem.
The Ackley function is a standard nonconvex test function characterized by a nearly flat outer region surrounding a steep, narrow global well:
\begin{align}
  f(\mathbf{x}) = & -20 \exp\!\left(-0.2 \sqrt{\frac{1}{d} \sum_{i=1}^d x_i^2}\right) \nonumber \\
               & - \exp\!\left(\frac{1}{d} \sum_{i=1}^d \cos(2\pi x_i)\right)
               + 20 + e,
\end{align}
where $d = 2$ in our experiments.
Although the cosine terms create many shallow undulations on the surface, the function has a single deep global well, and gradient-based methods from virtually any starting point converge to this well. The Ackley function is therefore not a test of global search; rather, it tests the \emph{precision} with which a solver can locate the minimum in a landscape where the steep, narrow well makes high accuracy difficult to achieve.

To avoid potential bias from QHD's uniform initialization on a symmetric domain, which could trivially favor origin-centered minima, we apply a deterministic random shift to the function. With seed 123, the global minimum is displaced to $\mathbf{x}^* \approx (0.962, 0.370)$, with $f(\mathbf{x}^*) = 0$. The computational domain is $[-5, 5]^2$.

We use a $32 \times 32$ grid (1,024 points) and vary the number of adaptive zoom iterations across $Z \in \{1, 7, 13, 19\}$, where $Z = 1$ corresponds to a single QHD run without any refinement. The classical baseline is 100-start L-BFGS-B with random initial points drawn uniformly from the domain. The best classical result across all 100 starts is reported.

Table~\ref{tab:ackley} summarizes the results. Without iterative refinement ($Z = 1$), QHD finds a solution with objective value $\sim 2.7 \times 10^{-1}$ and position error $\sim 6.2 \times 10^{-2}$ relative to the known minimum, consistent with the grid spacing $\Delta x = 10/32 \approx 0.31$, limiting accuracy to within one grid cell. As the iterations increase, QHD progressively narrows the search domain while maintaining the same $32 \times 32$ resolution, and both the objective value and position error decrease exponentially.

At $Z = 7$, the QHD objective value reaches $9.6 \times 10^{-6}$, already three orders of magnitude below the initial run. At $Z = 13$, QHD attains $2.2 \times 10^{-10}$, surpassing the 100-start L-BFGS-B baseline of $4.2 \times 10^{-9}$ in objective value. At $Z = 19$, the QHD objective reaches $4.0 \times 10^{-15}$, which is near the limit of double-precision floating-point arithmetic, with a corresponding position error of $9.2 \times 10^{-16}$, essentially machine epsilon.

These near-machine-precision numbers should be interpreted carefully. The Ackley study is a numerical stress test of the iterative refinement procedure, not a claim that downstream engineering applications require $10^{-15}$ accuracy. In practical optimization problems, such as the problems in power systems, tolerances on the order of $10^{-6}$ are often sufficient. This benchmark is to demonstrate that the iterative refinement can drive the error far below the coarse-grid spacing when high numerical accuracy is desired, while keeping the per-zoom qubit count fixed.

It is important to note that the 100-start L-BFGS-B baseline also locates the correct basin. Its best solution has position error $\sim 1.5 \times 10^{-9}$ relative to the known minimum. On the Ackley function, which is effectively unimodal for gradient-based methods, both QHD and L-BFGS-B find the global minimum, while the difference lies in precision. The steep and narrow Ackley well amplifies small position errors into non-negligible objective errors, so L-BFGS-B's $\sim 1.5 \times 10^{-9}$ position error translates to an objective of $4.2 \times 10^{-9}$, while QHD's iterative refinement reduces the position error by a further seven orders of magnitude. This comparison therefore demonstrates the \emph{precision} capability of iterative refinement, not a failure of the classical solver to identify the global basin.

\begin{table*}[t]
  \centering
  \caption{QHD iterative refinement on the shifted Ackley function ($32 \times 32$ grid, $N_t = 50{,}000$). The classical baseline is 100-start L-BFGS-B, which locates the correct basin with position error $\sim 1.5 \times 10^{-9}$ and best objective $4.17 \times 10^{-9}$. Bold entries indicate QHD achieves a lower objective than the classical baseline.}
  \label{tab:ackley}
  \smallskip
  \begin{tabular}{@{}ccccc@{}}
    \toprule
    Zoom $Z$ & QHD Objective & Position Error & Final $\Delta x$ & Wall Time (s) \\
    \midrule
    1  & $2.74 \times 10^{-1}$   & $6.19 \times 10^{-2}$  & $3.13 \times 10^{-1}$ & 10.6 \\
    7  & $9.59 \times 10^{-6}$   & $3.39 \times 10^{-6}$  & $1.45 \times 10^{-5}$ & 51.9 \\
    13 & $\mathbf{2.21 \times 10^{-10}}$ & $7.82 \times 10^{-11}$ & $< 10^{-10}$  & 139.9 \\
    19 & $\mathbf{4.00 \times 10^{-15}}$ & $9.16 \times 10^{-16}$ & $< 10^{-15}$  & 156.7 \\
    \midrule
    \multicolumn{2}{@{}l}{Classical (100-start L-BFGS-B)} & \multicolumn{3}{l}{$f = 4.17 \times 10^{-9}$} \\
    \bottomrule
  \end{tabular}
\end{table*}

\begin{figure}[htbp]
  \centering
  \subfloat[]{
  \includegraphics[width=0.85 \linewidth]{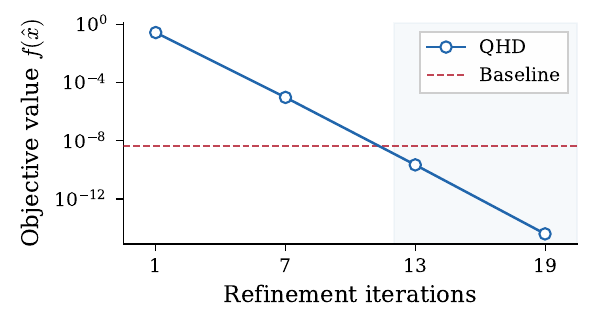}} \\
  \subfloat[]{
  \includegraphics[width=0.85 \linewidth]{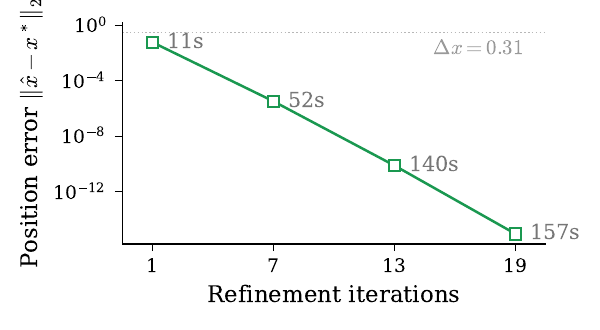}}
  \caption{Iterative refinement on the shifted Ackley function ($32 \times 32$ grid, $N_t = 50{,}000$, $x^* = (0.96, 0.37)$). (a)~QHD objective value vs.\ refinement iterations. The objective function values for QHD with refinement are shown using blue circle markers. The red dashed line marks the 100-start L-BFGS-B baseline. QHD's objective improves exponentially with zoom iterations and falls below the classical baseline at $Z = 13$. (b)~Position error $\|\hat{\mathbf{x}} - \mathbf{x}^*\|_2$ vs.\ refinement iterations, with wall-clock times annotated; the dotted line shows the grid resolution limit without zoom.}
  \label{fig:ackley_refinement}
\end{figure}

Figure~\ref{fig:ackley_refinement} visualizes the convergence behavior. Fig.~\ref{fig:ackley_refinement}a shows the QHD objective value decreasing with refinement iterations, with the classical baseline indicated by a dashed horizontal line. QHD crosses below the classical baseline between $Z = 7$ and $Z = 13$ and continues to improve. Fig.~\ref{fig:ackley_refinement}b shows the corresponding position error, annotated with the wall-clock time at each zoom level. The initial grid resolution limit $\Delta x \approx 0.31$ is shown as a dotted reference line. All zoom runs beyond $Z = 1$ achieve accuracy well below this limit, confirming that iterative refinement effectively decouples solution accuracy from the fixed grid size.

These results demonstrate that even on a very coarse $32 \times 32$ grid, the adaptive iterative refinement mechanism enables QHD to recover solutions at very high numerical precision. Since each iteration reuses the same grid size, the per-iteration computational cost (in terms of qubits and gates on quantum hardware) remains constant. The total cost scales linearly with the number of zoom iterations, making this approach particularly attractive for near-term quantum implementations where qubit counts are severely limited.

\textbf{Remark on wall-clock time: }
The wall-clock times reported in Table~\ref{tab:ackley} reflect the cost of \emph{classical simulation} of the QHD dynamics via the split-step Fourier method ($50{,}000$ Trotter steps per QHD run). The 100-start L-BFGS-B baseline completes in under one second on the same hardware. This asymmetry is expected. The purpose of classical simulation is to assess QHD's solution quality, not to compete with classical solvers on execution speed. On quantum hardware, each QHD run would correspond to a single Hamiltonian simulation circuit, and the relevant cost metric is circuit depth, qubit count, and quantum computer runtime, not classical simulation wall time. The wall-clock times are reported to characterize the cost of the zoom refinement procedure itself (linear scaling with $Z$), not to claim a computational advantage over classical methods on the Ackley function.

\subsection{Constrained benchmark: scaled Rastrigin function with ALM}
\label{sec:exp_rastrigin}
To evaluate the AL-QHD framework on constrained optimization, we use the scaled Rastrigin function with scaling factor $\alpha = 3$:
\begin{equation}
  f(\mathbf{x}) = 10d + \sum_{i=1}^d \left[ (3x_i)^2 - 10\cos(2\pi \cdot 3x_i) \right],
  \label{eq:scaled_rastrigin}
\end{equation}
with $d = 2$. With $\alpha = 3$, the oscillation frequency is tripled relative to the standard Rastrigin function, producing approximately 900 local minima on the domain $[-5, 5]^2$. This dense multimodal landscape is deliberately chosen to stress-test both QHD's global-search ability and the ALM's constraint-handling mechanism.

The function is separable: $f(x_0, x_1) = g(x_0) + g(x_1)$, where $g(t) = 10 + 9t^2 - 10\cos(6\pi t)$. The local minima of $g$ occur near $t \approx k/3$ for integer $k$, with function values approximately $g(k/3) \approx k^2$.

To avoid using constraints that can be handled by QHD through simple rectangular domain truncation, we impose nonlinear curved lower-bound constraints
\begin{align}
  x_0 &\geq 0.5 + 0.020(x_1-0.5)^2, \nonumber \\
  x_1 &\geq 0.5 + 0.020(x_0-0.5)^2 .
  \label{eq:curved_rastrigin_constraints}
\end{align}
This feasible set is non-rectangular. It excludes the unconstrained global minimum at the origin and the lower $k=1$ scaled-Rastrigin basin, while keeping the first feasible low-objective basin near $t^* \approx 0.6633$ in both coordinates. The known constrained optimum is therefore at $\mathbf{x}^* \approx (0.6633,0.6633)$ with $f(\mathbf{x}^*) \approx 7.960$.

QHD is simulated on a $64 \times 64$ grid (4,096 points) with $N_t = 50{,}000$ time steps. The ALM outer loop uses initial penalty $\rho_0 = 1$, doubling factor $\gamma = 2$, maximum penalty $\rho_{\max} = 10^9$, maximum 15 ALM iterations, and constraint tolerance $10^{-9}$. We vary the refinement iterations across $Z \in \{1, 2, 3, 4\}$. The main classical baseline is 100-start SLSQP (SciPy) with the same constraint specification, reporting the best feasible solution. We also perform a restart-budget sensitivity check for SLSQP by varying only the number of random starts while keeping the problem definition, stopping tolerance, and random-start generator fixed.

Table~\ref{tab:rastrigin} presents the results. The ALM outer loop converges with zero constraint violation in all refinement settings. The coarser $Z=1,2$ runs converge in 7 ALM iterations with final penalty parameter $\rho=64$, while the $Z=3,4$ runs converge in 6 ALM iterations with final penalty parameter $\rho=32$. This confirms that ALM handles the nonlinear inequalities without requiring extreme penalty values.

The main 100-start SLSQP baseline converges to a feasible point at approximately $(0.995,0.663)$ with objective value $12.93$. This corresponds to the classical solver settling in a mixed basin, with one coordinate in the $k=3$ basin and the other near the $k=2$ basin, yielding a significantly worse objective than the known constrained optimum.

\begin{table*}[ht]
  \centering
  \caption{AL-QHD on the scaled Rastrigin function ($\alpha = 3$, $64 \times 64$ grid, $N_t = 50{,}000$) with the nonlinear constraints in Eq.~\eqref{eq:curved_rastrigin_constraints}. The known constrained optimum is $f^* \approx 7.960$ at $(0.6633,0.6633)$. The 100-start SLSQP baseline converges to a different local basin at $(0.995,0.663)$ with $f = 12.93$. Bold entries indicate that the reported AL-QHD objective is lower than this budgeted SLSQP baseline.}
  \label{tab:rastrigin}
  \smallskip
  \begin{tabular}{@{}ccccccc@{}}
    \toprule
    Zoom $Z$ & QHD Objective & Err vs Known & ALM Iters & Cnstr Viol & $\rho_{\text{final}}$ & Time (s) \\
    \midrule
    1 & $\mathbf{12.89}$     & $5.42 \times 10^{-2}$ & 7 & 0 & 64 & 138.8 \\
    2 & $\mathbf{10.65}$     & $3.91 \times 10^{-2}$ & 7 & 0 & 64 & 262.0 \\
    3 & $\mathbf{8.089}$     & $8.52 \times 10^{-3}$ & 6 & 0 & 32 & 330.8 \\
    4 & $\mathbf{8.089}$     & $8.52 \times 10^{-3}$ & 6 & 0 & 32 & 364.1 \\
    \midrule
    \multicolumn{2}{@{}l}{Classical (100-start SLSQP)} & \multicolumn{5}{l}{$f = 12.93$ at $(0.995,0.663)$} \\
    \bottomrule
  \end{tabular}
\end{table*}

In this budgeted comparison, AL-QHD returns a lower objective than the 100-start SLSQP baseline at every zoom level. Even without refinement ($Z=1$), QHD achieves an objective $12.89$, marginally better than the SLSQP result of $12.93$. Although the improvement is small in absolute terms ($<0.4\%$), the two solutions lie in qualitatively different regions. QHD's solution at $(0.625,0.625)$ is in the correct basin near the $k=2$ minimum in both coordinates, limited by grid resolution, while SLSQP's solution places one coordinate in the higher-energy $k=3$ basin. This indicates that AL-QHD can identify the correct constrained basin even at coarse resolution.

With two refinement iterations, QHD improves to $10.65$, representing an $18\%$ lower objective than the main classical baseline. At $Z=3$ and $Z=4$, AL-QHD returns a feasible point with objective $8.089$ and position error $8.52\times10^{-3}$ relative to the known optimum, corresponding to a $1.63\%$ relative objective excess. The 100-start classical baseline, by contrast, is trapped in a local basin with relative objective excess $(f_{\mathrm{SLSQP}} - f^*)/f^* = (12.93 - 7.96)/7.96 \approx 62.5\%$. Thus, this nonlinear constrained example supports a softened conclusion: AL-QHD finds a near-optimal feasible point in the correct basin under a moderate restart-budget comparison, rather than exactly reaching the true constrained optimum in the current run.

\begin{figure}[ht]
  \centering
  \subfloat[]{\includegraphics[width=0.85\linewidth]{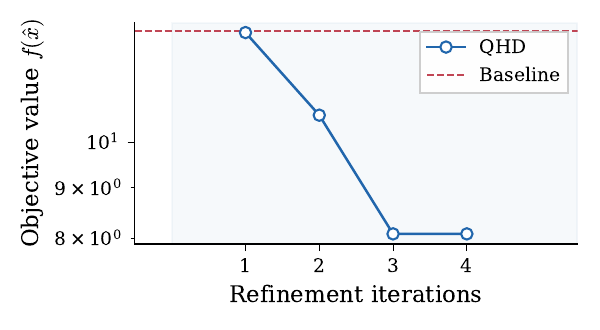}} \\
  \subfloat[]{\includegraphics[width=0.85\linewidth]{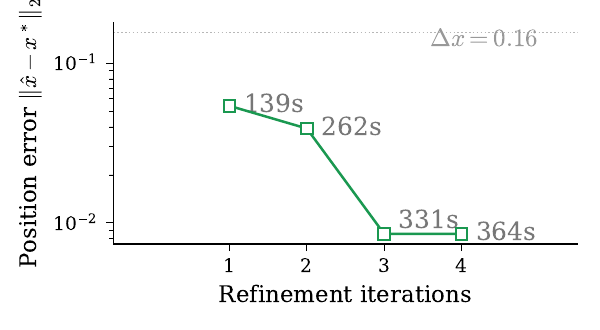}}
  \caption{AL-QHD on the nonlinearly constrained scaled Rastrigin function ($\alpha = 3$, $64 \times 64$ grid, $N_t = 50{,}000$), with $x^* = (0.66, 0.66)$. (a)~QHD objective value vs.\ refinement iterations. The objective function values for QHD with refinement are shown using blue circle markers. The red dashed line marks the 100-start SLSQP baseline. AL-QHD identifies the correct constrained basin at coarse resolution and reaches a near-optimal feasible point by $Z=3$. (b)~Position error $\|\hat{\mathbf{x}} - \mathbf{x}^*\|_2$, annotated with wall-clock times. The dashed line represents the positional error produced by the baseline solution.}
  \label{fig:rastrigin_refinement}
\end{figure}

Fig.~\ref{fig:rastrigin_refinement} illustrates the convergence behavior. The key improvement occurs between $Z=2$ and $Z=3$, where QHD moves from the objective level $10.65$ to $8.089$ and approaches the known constrained optimum. The $Z=4$ run does not further improve the best feasible solution, which reflects the current ALM stopping rule and finite-grid refinement behavior rather than a change in feasibility.

\paragraph{Budgeted-baseline analysis for the scaled Rastrigin benchmark:}
The behavior of the 100-start SLSQP baseline is instructive. The scaled Rastrigin function with $\alpha=3$ places low local minima near $t\in\{0.6633,0.995,1.328,\dots\}$ in each coordinate, and the nonlinear constraints in Eq.~\eqref{eq:curved_rastrigin_constraints} preserve the same relevant low-objective basins while making the feasible set non-rectangular. Since SLSQP is a local method, each random start converges to a nearby feasible local minimum. With 100 starting points drawn uniformly from $[-5,5]^2$, the best run finds a mixed basin with one coordinate near $0.995$ and the other near $0.663$, yielding a substantially higher objective than the known optimum.
QHD, by contrast, evolves a wavefunction that explores the entire domain simultaneously inside each ALM subproblem. The wavefunction-based search can place probability mass in the correct low-objective basin, while the ALM outer loop drives feasibility for the nonlinear constraints. The iterative zoom then refines this coarse localization at a fixed per-run grid size.

\paragraph{Sensitivity to the classical restart budget:}
Table~\ref{tab:rastrigin_restart_budget} shows how the best feasible SLSQP solution changes when the number of random starts is varied while keeping the same objective, constraints, tolerance, and random-start generator. The dependence on restart budget is substantial: with 10, 20, 50, and 100 starts, the best classical solution remains in a mixed basin; by 200 starts, SLSQP recovers the known constrained optimum, and 500 starts also succeeds. We therefore interpret the main constrained comparison as a comparison against a moderate multi-start local-search budget, not as evidence that restart-based classical methods can never recover the optimum.

\begin{table*}[t]
  \caption{Sensitivity of the scaled-Rastrigin SLSQP baseline to the number of random starts under the nonlinear constraints in Eq.~\eqref{eq:curved_rastrigin_constraints}. All runs use the same problem definition, stopping tolerance, and random-start generator as the main 100-start baseline; only the restart budget changes. The distance column reports Euclidean distance to the known optimum $x^*\approx(0.6633,0.6633)$.}
  \label{tab:rastrigin_restart_budget}
  \smallskip
  \centering
  \begin{tabular}{@{}cccc@{}}
    \toprule
    SLSQP starts & Best feasible point & Best feasible objective & Distance to $x^*$ \\
    \midrule
    10  & $(0.995, 0.663)$ & $12.93$ & $0.332$ \\
    20  & $(0.663, 0.995)$ & $12.93$ & $0.332$ \\
    50  & $(0.663, 0.995)$ & $12.93$ & $0.332$ \\
    100 & $(0.995, 0.663)$ & $12.93$ & $0.332$ \\
    200 & $(0.663, 0.663)$ & $7.95966$ & $<10^{-3}$ \\
    500 & $(0.663, 0.663)$ & $7.95966$ & $<10^{-3}$ \\
    \bottomrule
  \end{tabular}
\end{table*}

\subsection{Summary of benchmark findings}
The two benchmarks yield complementary insights about different aspects of QHD.
On the unconstrained Ackley function, both QHD and 100-start L-BFGS-B successfully locate the global minimum's basin. The primary finding is that QHD's iterative refinement can overcome the fundamental resolution limit of a fixed grid, achieving near-machine-precision accuracy ($\sim 4 \times 10^{-15}$ in objective) from a $32 \times 32$ grid, which provides a qubit-efficient approach relevant to near-term quantum hardware. This Ackley result can be viewed as a stress test of refinement accuracy, rather than claiming machine-precision solutions for other downstream applications.
The main point is that the per-iteration qubit and gate cost remains fixed while the effective spatial resolution improves exponentially with the number of zoom iterations.

On the nonlinearly constrained scaled Rastrigin function, we find that the AL-QHD framework can identify the correct constrained basin and return a near-optimal feasible point on a landscape with approximately 900 local minima, whereas the 100-start SLSQP baseline used for the main comparison remains trapped in a suboptimal mixed basin. This result should be interpreted as evidence that the AL-QHD workflow can function on a dense multimodal nonlinear constrained benchmark under a fixed restart-budget comparison, not as a general proof of superiority over classical methods. Even at coarse resolution ($Z=1$), QHD identifies the correct basin in both coordinates, while the 100-start classical baseline does not. The restart-budget sensitivity study shows that larger multi-start budgets improve the classical result and recover the optimum by 200 starts, so the comparison should be read as a budgeted baseline comparison rather than an impossibility claim against classical restart methods. The ALM framework provides effective nonlinear constraint handling with moderate penalty parameters, while the QHD inner solver supplies a wavefunction-based global search mechanism for the unconstrained ALM subproblems.

\subsection{Practical resource needs of AL-QHD: Power-system case study}
\label{sec:power}

To understand the resources required to use the AL-QHD framework for practical constrained optimization, we use the ACOPF problem from power systems as one representative example for exercising the gate-level cost-analysis pipeline.
Specifically, we use ACOPF-derived symbolic models to generate realistic nonconvex constrained instances, map them into the AL-QHD workflow, and estimate gate costs under two hardware-oriented views (NISQ and FTQC). Related quantum power-flow work has explored HHL-based sparse linear-system solvers and variational circuits, while also highlighting QRAM, circuit-depth, and noise/resource barriers~\cite{liu2024quantumpower, a18080491}. The purpose of this case study is to stress-test the resource scaling of AL-QHD on a structured constrained optimization model.

\subsubsection{AC optimal power flow}
ACOPF is a fundamental nonconvex optimization problem in power systems that seeks a minimum-cost feasible steady-state operating point while enforcing nonlinear AC power-flow equations and practical operating limits, including generator, voltage, and line-flow constraints~\cite{molzahn2019survey}.
In a typical ACOPF formulation, decision variables include bus voltage magnitudes $V_i$, voltage angles $\theta_i$,
and generator outputs $(P_{G,i},Q_{G,i})$.
The objective usually minimizes generation cost, commonly modeled as a quadratic function
\begin{equation}
  \sum_{i\in\mathcal{G}} C_i(P_{G,i}) = \sum_{i\in\mathcal{G}} \left(a_i P_{G,i}^2 + b_i P_{G,i} + c_i\right),
\end{equation}
subject to:
(i) AC power balance equations $P_i(V,\theta),Q_i(V,\theta)$ at each bus,
(ii) generator capacity limits,
(iii) voltage limits, and
(iv) optionally line-flow limits.
The nonconvexity stems primarily from products and trigonometric terms:
$V_iV_j\cos(\theta_i-\theta_j)$ and $V_iV_j\sin(\theta_i-\theta_j)$.

\textbf{How AL-QHD is applied:}
We apply ALM to the equality constraints (AC power balance and reference-angle constraints), and treat box constraints (voltage and generator bounds) directly via domain truncation and/or ALM variants. The inner QHD solver is tasked with minimizing the augmented Lagrangian subproblem, which is itself nonconvex.

\subsubsection{Problem setup: Power-system instances}

For this case study, we focus on connected subgraphs extracted from the Texas7k synthetic transmission grid~\cite{texas7k,holzer2024grid} 
For each target problem size, the scripts expand a connected neighborhood from the transmission grid, build the associated ACOPF-derived symbolic model, and retain the realized number of active variables after subgraph extraction and simplification.
The extracted instances in this study range from 2 to 198 buses, and the final one-hot Hamiltonians span from 6 to 538 active variables. The more detailed extraction procedure is listed in Appendix~\ref{app:acopf}.

There is no single fixed bus-to-variable conversion. The realized active-variable count depends on which buses, generators, and constraints remain active after extraction and symbolic simplification. Table~\ref{tab:bus_to_variables} gives representative examples together with the realized active-variables-per-bus ratio. This ratio can vary substantially across power-system configurations and should not be interpreted as a standard practical metric for power-system analysis. Here, it is used only to provide a characterization of the ACOPF problem set used in this work.
As these experiments are intended to characterize quantum resource scaling on realistic nonconvex power-system instances, rather than to claim end-to-end large-scale OPF solution quality, the metrics reported here emphasize qubit and gate scaling.

\begin{table}[ht]
  \centering
  \caption{Representative mapping from realized Texas7k subgraph size to active-variable count in the ACOPF-derived symbolic model. The conversion is not fixed; it depends on the extracted neighborhood and the simplifications that remain active.}
  \label{tab:bus_to_variables}
  \smallskip
  \begin{ruledtabular}
  \begin{tabular}{@{}ccc@{}}
    \textbf{Buses}  & \textbf{Active variables} & \textbf{Variables per bus} \\
    \midrule
    2   & 6   & 3.00 \\
    5   & 14  & 2.80 \\
    12  & 34  & 2.83 \\
    32  & 104 & 3.25 \\
    70  & 198 & 2.83 \\
    132 & 388 & 2.94 \\
    198 & 538 & 2.72 \\
  \end{tabular}
  \end{ruledtabular}
\end{table}

\subsubsection{Resource estimation results}

\begin{figure}[htbp]
  \centering
   \subfloat[NISQ Cost: Hard vs Easy Gates]{\includegraphics[width=0.85 \linewidth]{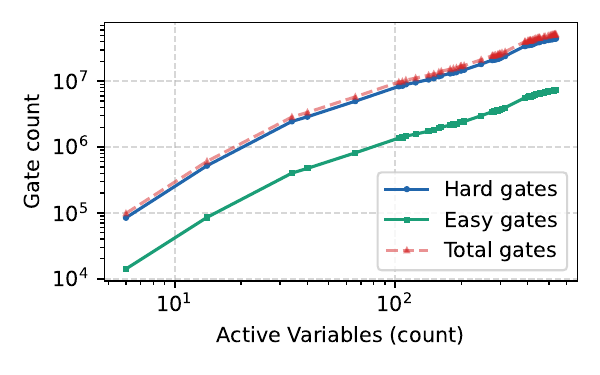}}

  \medskip
    \subfloat[FT Cost: T-Count vs Clifford]{\includegraphics[width=0.85 \linewidth]{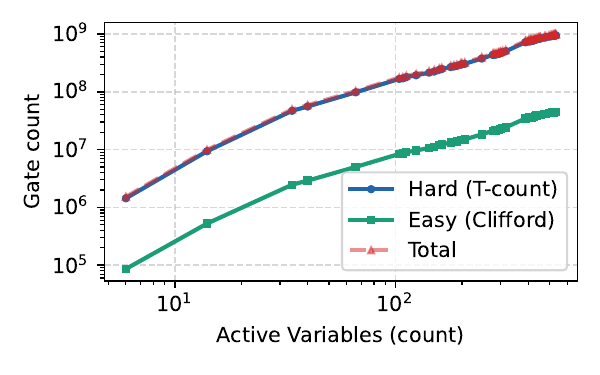}}

    \subfloat[FT T-Count Breakdown]{\includegraphics[width = 0.85 \linewidth]{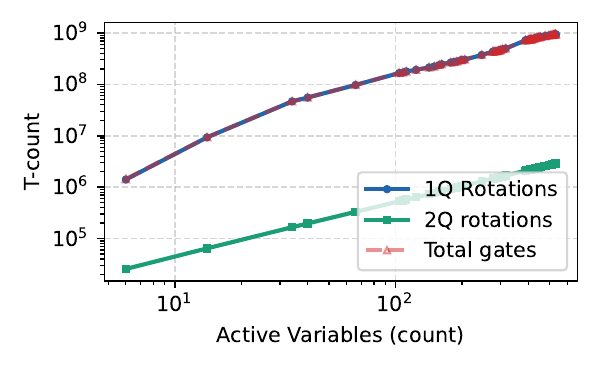}}
  \caption{Gate count scaling for one-hot encoded QHD Hamiltonians derived from Texas7k ACOPF subgraphs. (a) report the NISQ-oriented model, where the red dashed curve is the total gate count, the blue curve is the hard two-qubit entangling-gate count, and the teal curve is the easy single-qubit gate count. Most of the gates are entangling gates in this case. (b-c) report the FTQC-oriented model, where in (b) the red dashed curve is the total gate count, the blue curve is the hard T-gate count from exact and synthesized rotations, and the teal curve is the easy Clifford count. In (c), we show the breakdown of the T gates. The blue curve shows the number of T gates synthesized from single-qubit rotations, while the teal curve shows the T gate counts from the two-qubit rotations. The total T count is plotted as the red dashed curve. In both cases, the hard resource grows more rapidly than the easy resource and dominates the total gate count on the larger extracted instances.}
  \label{fig:gate_scaling}
\end{figure}

Fig.~\ref{fig:gate_scaling} shows the estimated gate-level resource scaling for the ACOPF instances. As discussed in Sec.~\ref{sec:meth:res_model}, we consider both NISQ-oriented and FTQC-oriented resource analyses.
In the NISQ era, the dominant cost is from the entangling gates, which grows from approximately $8.42\times 10^4$ to $4.46\times 10^7$ gates across the extracted instances, while easy single-qubit gates remain smaller by roughly one order of magnitude.
In the FTQC era, the dominant cost is from the synthesized-rotation T gates, which grows from approximately 
$1.43\times 10^6$ to $9.42\times 10^8$, whereas the accompanying Clifford overhead grows from $8.61\times 10^4$ to $4.46\times 10^7$.
The fault-tolerant curve therefore reinforces the same qualitative conclusion as the NISQ curve: resource growth is driven by the hard part of the Hamiltonian simulation, and the gap between hard and easy resources widens substantially as the number of active variables increases.
This scaling motivates both lower-resolution encodings and iterative-refinement strategies that keep the qubit footprint fixed while improving solution quality across multiple QHD runs.

The Texas7k case study shows that realistic symbolic power-system Hamiltonians rapidly become hard-gate dominated in both NISQ and FTQC implementations. For NISQ devices, two-qubit entangling gates dominate the circuit cost and grow to approximately $4.46\times 10^7$ on the largest extracted instance. For fault-tolerant devices, the corresponding T-count reaches approximately 
$9.42\times 10^8$, with Clifford overhead remaining substantially smaller.
The resource analysis should be interpreted as an implementation-level estimate rather than an algorithmic complexity bound. It nevertheless makes the practical bottlenecks clear: without additional structure exploitation, synthesized rotations and entangling operations dominate the cost of gate-based AL-QHD workflow on realistic constrained power-system models. Thus, if AL-QHD is to become relevant for ACOPF-scale applications, the present estimates point toward fault-tolerant rather than near-term hardware.

The saved resource breakdown shows that the potential evolution, rather than the kinetic template, is the leading hard-cost source for these ACOPF-derived instances. In the NISQ model, the compiled kinetic template contributes only $44n$ hard two-qubit gates for $n$ active variables; the potential parity-decomposition block accounts for $99.7\%$ of the hard gates at $n=6$ and $99.95\%$ at $n=538$. In the FTQC model, the compiled kinetic template contributes $42n$ generic rotations plus $2n$ exact $T$ gates, while the potential block contributes millions of synthesized $R_z$ rotations on the largest instances. The potential-side contribution accounts for $98.2\%$ of the hard $T$ count at $n=6$ and $99.7\%$ at $n=534$.
A log--log fit of the saved sampled FTQC hard counts gives an empirical scaling 
$T_{\mathrm{hard}}\approx 3.48\times 10^5 n^{1.27}$ over the extracted instances ($R^2\approx 0.98$ in log space).
The analogous NISQ hard-count fit is approximately $2.17\times 10^4 n^{1.23}$ ($R^2\approx 0.98$). These results also indicate that the present quantum resource-estimation framework remains limited to small ACOPF subgraphs or small IEEE test cases, and is not yet scalable to full-size classical OPF benchmarks. 

\section{Discussion}
\label{sec:discussion}

The experimental results highlight several findings relevant to the practical deployment of QHD for nonconvex optimization.

\paragraph{Iterative refinement as a qubit-saving strategy.}
The Ackley benchmark demonstrates that iterative refinement fundamentally changes the resource-accuracy tradeoff for QHD. On quantum hardware, the number of qubits scales with the grid resolution $r$ (specifically, $d \cdot r$ qubits under one-hot encoding for $d$ variables at resolution $r$). Without refinement, achieving position accuracy $\epsilon$ requires $r \sim 1/\epsilon$, leading to prohibitive qubit counts for high-precision applications. With iterative refinement, a fixed, small $r$ suffices: the $32 \times 32$ grid used in the Ackley experiment requires only $2 \times 32 = 64$ qubits under one-hot encoding, yet achieves $10^{-15}$ accuracy through 19 zoom iterations. Each zoom iteration requires one additional QHD circuit execution at the same qubit count, so the cost scales linearly in zoom iterations rather than exponentially in precision. We emphasize that on the Ackley function, the classical solver L-BFGS-B also finds the correct basin with high positional accuracy ($\sim 1.5 \times 10^{-9}$). The comparison here does not demonstrate a fundamental quantum advantage in solution quality, but rather validates iterative refinement as a viable precision-enhancement mechanism for QHD that keeps the qubit count fixed. It is also not a claim that power-system applications require machine-precision answers; instead, it shows that QHD refinement can exceed the coarse-grid resolution by many orders of magnitude when a benchmark demands it.

\paragraph{AL-QHD behavior on multimodal constrained landscapes.}
The nonlinearly constrained scaled Rastrigin benchmark demonstrates a qualitatively different behavior from the Ackley precision test. QHD's wavefunction-based exploration, combined with the ALM outer loop, enables AL-QHD to identify the correct constrained basin on a densely multimodal landscape where the 100-start SLSQP baseline used for the main comparison remains trapped in a suboptimal mixed basin. Notably, even at the coarsest resolution ($Z=1$, no iterative refinement), QHD already identifies the correct basin in both coordinates, while none of the 100 classical starts do so in both coordinates simultaneously. At the same time, the restart-budget sensitivity study shows that this comparison is budget-dependent. Larger classical multi-start budgets do improve the SLSQP result and recover the optimum by 200 starts.

However, whether this behavior persists in higher dimensions is an important open question. On the one hand, the number of local minima grows combinatorially with dimension, and the probability that any single classical start lands in the global basin shrinks rapidly. On the other hand, the classical simulation cost of QHD also grows exponentially with dimension, and the behavior of the QHD wavefunction on high-dimensional landscapes has not been extensively studied due to the limited classical simulation capabilities. We regard this as an important direction for future work.

\paragraph{ALM as a practical constraint-handling framework.}
The ALM outer loop converges reliably with moderate penalty parameters ($\rho_{\text{final}}\in\{32,64\}$ in the Rastrigin experiments), confirming that ALM provides an effective interface between the unconstrained QHD solver and nonlinear constrained optimization problems. The constraint violation is zero in all reported Rastrigin runs, and ALM convergence requires only 6--7 iterations. This is consistent with the theoretical motivation, i.e., by using Lagrange multipliers to track constraint satisfaction, ALM avoids the need for extreme penalty values that would distort the potential energy landscape.

\paragraph{Limitations and future work.}
Several limitations should be noted. First, our benchmarks are two-dimensional; scaling to the higher-dimensional problems encountered in power systems (tens to hundreds of variables) will require careful attention to qubit overhead and Trotter error accumulation. Second, the classical simulation cost of the split-step Fourier method grows exponentially with dimension, so classical benchmarking of QHD is limited to low-dimensional problems. Third, the iterative refinement strategy assumes that QHD correctly identifies the basin of the global minimum at coarse resolution; on sufficiently complex landscapes, this assumption may fail, and multiple coarse runs or alternative initialization strategies may be needed. Fourth, our classical baselines are still limited in scope: the constrained comparison is sensitive to restart budget, and the power-system case study does not include end-to-end ACOPF solver comparisons against mature power-system baselines. In addition, the present analysis focuses mainly on the Trotterized realization used in Ref.~\cite{leng2023quantum}. 

Potentially more scaling-friendly and resource-efficient implementations for large-scale optimization problems, such as those based on the quantum Fourier transform or more advanced Quantum Hamiltonian simulation techniques, are not considered here.
Recent fault-tolerant resource studies for quantum linear-system algorithms show that assessing practical advantage requires accounting not only for asymptotic speedups, but also for physical qubits, runtime, energy, magic-state  distillation, and data-access overheads~\cite{tu2025towards}.
A more rigorous end-to-end resource analysis, incorporating QEC code-algorithm co-design, will be needed to obtain tighter resource estimates and to assess the problem scales at which quantum advantage may be plausible. Furthermore, more extensive classical baselines, including global solvers such as differential evolution, basin-hopping, branch-and-bound methods, and standard ACOPF workflows, would provide a more comprehensive comparison and are planned for future work.

\section{Conclusion}
\label{sec:conclusion}

This paper studies QHD within a hybrid AL-QHD framework for constrained nonconvex optimization, motivated by applications in power system operation. The main objective is to test whether the AL-QHD workflow can work on representative constrained examples and to estimate the cost of a digital implementation, rather than to establish a broad quantum-over-classical advantage. Our experiments on standard test functions and ACOPF-derived power-system resource models demonstrate three key findings.

First, iterative refinement enables QHD to achieve near-machine-precision accuracy ($\sim 4 \times 10^{-15}$ on the Ackley function) from a coarse $32 \times 32$ grid. On this essentially unimodal function, both QHD and 100-start L-BFGS-B correctly identify the global basin; the comparison validates iterative refinement as a mechanism for achieving high precision at fixed qubit cost, with the effective spatial resolution improving exponentially with the number of zoom iterations. This is best interpreted as a stress test of refinement accuracy rather than an application-level precision requirement.

Second, on a nonlinearly constrained scaled Rastrigin function with approximately 900 local minima, AL-QHD identifies the correct constrained basin and returns a near-optimal feasible point with objective $8.089$, which is $1.63\%$ above the known constrained optimum $f^*\approx7.96$. The 100-start SLSQP baseline used in the main comparison converges to a suboptimal mixed basin with relative objective excess $(f_{\mathrm{SLSQP}}-f^*)/f^*\approx62.5\%$. Even at coarse resolution, QHD identifies the correct basin in both coordinates, showing that the AL-QHD workflow can function on this multimodal nonlinear constrained benchmark under a moderate restart-budget comparison. A restart-budget sensitivity check shows that the classical result improves with more starts and reaches the optimum by 200 starts. The ALM outer loop converges reliably with moderate penalty parameters and zero constraint violation.

Third, the power-system resource study shows steep but interpretable implementation-level scaling: in our Texas7k-derived instances, the NISQ hard-gate count reaches approximately $4.46\times 10^7$ while the fault-tolerant T-count reaches approximately 
$9.42\times 10^8$ at around $5.3\times 10^2$ active variables, with hard gates dominating in both views.

These results support the viability of QHD as an inner solver for studying constrained nonconvex optimization, with iterative refinement providing a practical path to surpass the native coarse-grid resolution on qubit-limited hardware. However, applying AL-QHD to ACOPF at scales relevant for potential quantum advantage would likely require a large-scale fault-tolerant quantum computer. Important open questions remain regarding scalability to higher-dimensional problems, performance on power-system-scale instances, and tighter resource bounds for power-system optimization needed to assess whether any advantage in solution quality or runtime is plausible.

\section*{Acknowledgments}

The authors would like to thank Jiaqi Leng and Zhixin Song for the fruitful discussions.
This research was supported by Pacific Northwest National Laboratory’s Quantum Algorithms and Architecture for Domain Science (QuAADS) Laboratory Directed Research and Development (LDRD) Initiative.
This research used resources of the National Energy Research Scientific Computing Center (NERSC), a U.S. Department of Energy Office of Science User Facility located at Lawrence Berkeley National Laboratory, operated under Contract No. DE-AC02-05CH11231.
The Pacific Northwest National Laboratory is operated by Battelle for the U.S. Department of Energy under Contract DE-AC05-76RL01830.

\appendix

\section{FT rotation-synthesis resource overhead}
\label{app:ft_factor}

\begin{figure}[ht]
    \centering
    \includegraphics[width=0.9\linewidth]{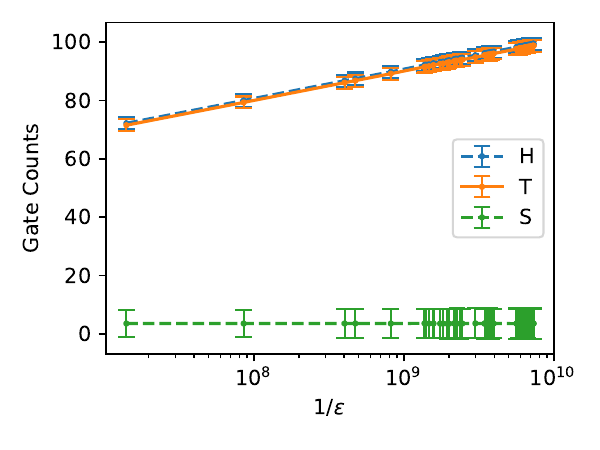}
    \caption{Average $H$ (blue), $S$ (green), and $T$ (orange) gate counts per synthesized rotation for approximating $R_z$ rotations using gridsynth. For each ACOPF case study, the per-rotation synthesis tolerance $\epsilon$ is determined from the total number of $R_z$ rotation gates and the target QHD-run accuracy.  At each tolerance, 5000 rotation angles are sampled uniformly from $(-\pi,\pi]$, and the corresponding $R_z(\theta)$ gates are synthesized using the gridsynth implementation in \texttt{NWQEC}. The error bars indicate the standard deviation over the 5000 sampled rotation angles.}
    \label{fig:gridsynth}
\end{figure}

Instead of synthesizing every $R_z$ rotation individually for each AL-QHD instance, we estimate the average $T$-gate overhead per $R_z$ rotation. Specifically, for each target approximation tolerance $\epsilon$, we uniformly sample 5000 random rotation angles from $(-\pi, \pi]$, and synthesize the corresponding $R_z(\theta)$ gates using the gridsynth method implemented in the \texttt{NWQEC} package~\cite{nwqec_toolkit, wang2024optimizing}. We define the synthesis factor $r(\epsilon)$ as the average number of $T$ gates required to approximate a single $R_z$ rotation to the accuracy target $\epsilon$.

When we are given a case study from AL-QHD implementation, in our estimation, we set the total accuracy of evaluating the constraint violation to be $10^{-3}$, while assuming the violation penalty factor $\rho \sim 10^3$. Therefore, each QHD run within the AL-QHD framework will require a target accuracy $10^{-6}$. From each case study, when we have the estimation of the total number of rotation gates $N_r = N_{r, K} + N_{r, V}$, each rotational gate approximation accuracy is estimated by $\epsilon \sim 10^{-6} / N_r$. For this case-specific tolerance, we compute $r(\epsilon_{\rm rot})$ using the 5000-sample gridsynth procedure described above.

Fig.~\ref{fig:gridsynth} shows the average gate counts obtained by synthesizing $R_z$ rotations with gridsynth, using 5000 uniformly sampled random rotation angles for each required synthesis tolerance $\epsilon$. Each data point corresponds to one ACOPF case study considered in the main text. Consistent with Ref.~\cite{gridsynth}, the synthesis factor $r$ grows approximately linearly with $\log(1/\epsilon)$. We use the case-specific values of $r$ obtained from these sampling experiments to estimate the total $T$-gate counts reported in Fig.~\ref{fig:gate_scaling}. Even though the number of generated $H$ gates also grows with the $T$ gates, as shown by Refs.~\cite{wang2024optimizing, wang2025tableau}, these gates can be removed with proper compilation optimizations.

\section{Power-system model generation} \label{app:acopf}
Our current pipeline
constructs AC power system subproblems as follows:
\begin{enumerate}[leftmargin=*, noitemsep]
  \item Load a MATPOWER case file of Texas7k synthetic grid data. 
  \item Select a connected subgraph of size \texttt{target\_n} buses by BFS expansion from a seed bus (typically chosen by generation capacity).
  \item Build the complex bus admittance matrix $Y_{\text{bus}}$ for the subgraph.
  \item Construct a symbolic ACOPF objective and constraints using standard AC power flow equations.
  \item Form a penalized objective by adding quadratic penalties for equality constraints.
  \item Encode the resulting objective via one-hot into an Ising Hamiltonian (diagonal Pauli-$Z$ strings).
\end{enumerate}
The same general workflow could in principle be adapted to related dispatch formulations, but the concrete power-system instances reported in this paper are ACOPF-derived.
\bibliography{bib}

\end{document}